\documentclass[twocolumn,nofootinbib,jcp,preprintnumbers,amsmath,amssymb]{revtex4-1}
\usepackage{natbib,hyperref}
\usepackage{amsmath}
\usepackage{graphicx}
\usepackage{dcolumn}
\usepackage{bm}
\usepackage{epsfig,color,xspace,multirow,xr,bbold}
\usepackage[all]{xy}
\usepackage{setspace}
\usepackage{url}
\usepackage[colorinlistoftodos]{todonotes}
\usepackage{threeparttable} 

\usepackage{natbib,hyperref}
\usepackage{float}
\usepackage{placeins}
\usepackage[utf8]{inputenc}

\begin{document}

\title{
Troubleshooting Unstable Molecules in Chemical Space
}
\date{\today}     

\author{Salini Senthil$^{1}$}
\author{Sabyasachi Chakraborty$^{1}$}
\author{Raghunathan Ramakrishnan$^{1}$}

\email{ramakrishnan@tifrh.res.in}

\affiliation{$^1$Tata Institute of Fundamental Research, Centre for Interdisciplinary Sciences, Hyderabad 500107, India}


\begin{abstract}
A key challenge in automated chemical compound space explorations
is ensuring veracity in minimum energy
geometries---to preserve intended bonding
connectivities. We discuss an iterative high-throughput workflow for
connectivity preserving geometry optimizations
exploiting the nearness between quantum mechanical models.
The methodology is benchmarked on the QM9 dataset
comprising DFT-level properties of 133,885 small molecules;
of which 3,054 have questionable geometric stability.
We successfully troubleshoot 2,988 molecules
and ensure a bijective mapping between desired Lewis formulae and final geometries.
Our workflow, based on
DFT and post-DFT methods, identifies 66 molecules as
unstable; 52 contain  $-{\rm NNO}-$, the rest are strained
due to pyramidal sp$^2$ C.
In the curated dataset, we inspect molecules with
long CC bonds and identify ultralong contestants
($r>1.70$~\AA{}) supported by
topological analysis of electron density.
We hope the proposed strategy to play a role 
in big data quantum chemistry initiatives.
\end{abstract}

\maketitle

The central focus of all explorations in the chemical compound 
space (CCS)---designed using mathematical graphs without any bias---is 
to establish property trends across it, to steer 
efforts towards synthesis and 
experimental characterization of molecules with desired properties\cite{kirkpatrick2004chemical}. 
Even a tiny fraction of CCS spanned by small organic molecules, is too
vast\cite{ruddigkeit2012enumeration} to exhaustively cover with the
conventional ``one-molecule-at-a-time'' simulations paradigm. 
To tackle such hard problems, 
high-throughput computation\cite{pyzer2015high,ramakrishnan2014quantum} combined with 
big data analytics\cite{tetko2016bigchem,gomez2020machine} and
machine learning techniques\cite{von2018quantum,von2020retrospective,gomez2020machine,sanchez2018inverse,tkatchenko2020machine}
promises novel strategies---influencing nearly 
all aspects of {\it ab initio} molecular modelling. 
At the inception of molecular big data,
it is crucial to ensure  
equilibrium geometries are
faithfully mapped to their desired Lewis formulae\cite{lewis1916atom,purser1999lewis,purser2001lewis}.

A previous high-throughput study presented density functional theory (DFT) level geometries and static properties of the QM9 dataset with
133,885 (134k) closed-shell organic molecules comprising up to nine of C, O, N and F atoms\cite{ramakrishnan2014quantum}.
As of yet, 2.3\% of this dataset---amounting to 3,054 (3k) molecules---remains 
``uncharacterized'', where the reported geometries are
not in line with the intended Lewis structures.
This structural ambiguity stems from the 
fact that the initial atomic coordinates, generated from SMILES (Simplified Molecular Input Line Entry System) descriptors---denoted
SMI, henceforth---have undergone rearrangements during DFT treatment 
due to their intrinsic geometric metastability\cite{lauderdale1992stability}.
Hence, the 3k subset was excluded from other
data-intensive explorations \cite{narayanan2019accurate,ramakrishnan2015big,christensen2020fchl,dandu2020quantum}. 
\begin{figure}[h]
    \centering
    \includegraphics[width=8.8cm]{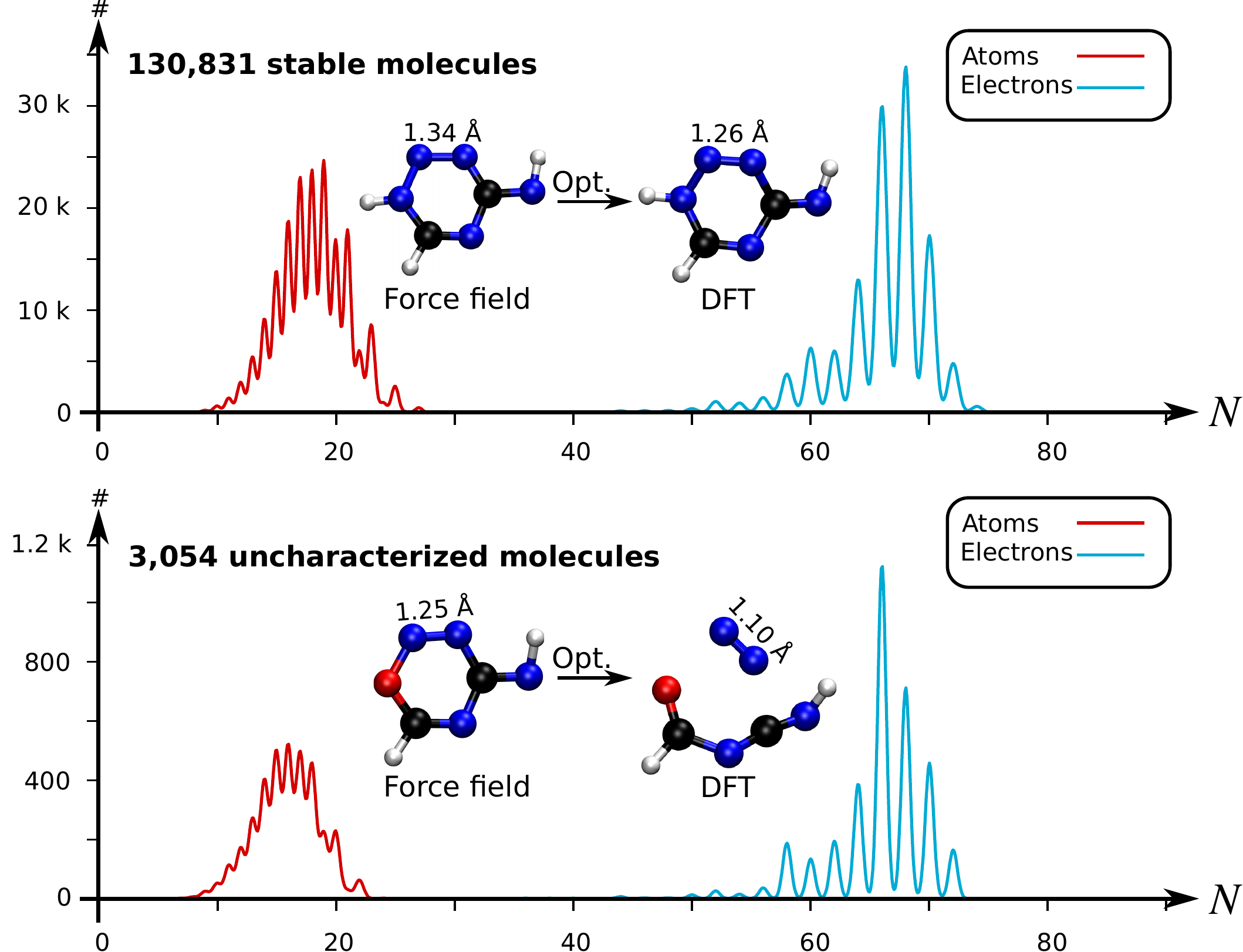}
    \caption{Distribution of atoms (red lines) and electrons (cyan lines) per molecule
    in the QM9 dataset: 
    Stable 131k (top) and uncharacterized 3k (bottom) subsets.
    For two similar molecules, across the sets, 
    the effect of DFT-level relaxation starting with a
    force field based geometry is shown.
    }
    \label{fig:distributionfig1}
\end{figure}

To capture potential deviations in chemical trends between
the stable 131k set and the 3k uncharacterized subset, we investigated 
their distributions, presented in Fig.~\ref{fig:distributionfig1}.
With both sets exhibiting surprisingly similar trends in molecular 
composition and size, the reason behind their contrast in stabilities
remains elusive. 
Our attempts to rationalize this
trend solely using molecular composition and functional groups bore no fruits, 
suggesting in turn the ``Anna Karenina Principle'' at work---all geometrically stable molecules are stabilized through common factors 
and each molecule failing to converge does so for a unique reason!

\begin{figure*}[!hptb]
    \centering
    \includegraphics[width=16cm]{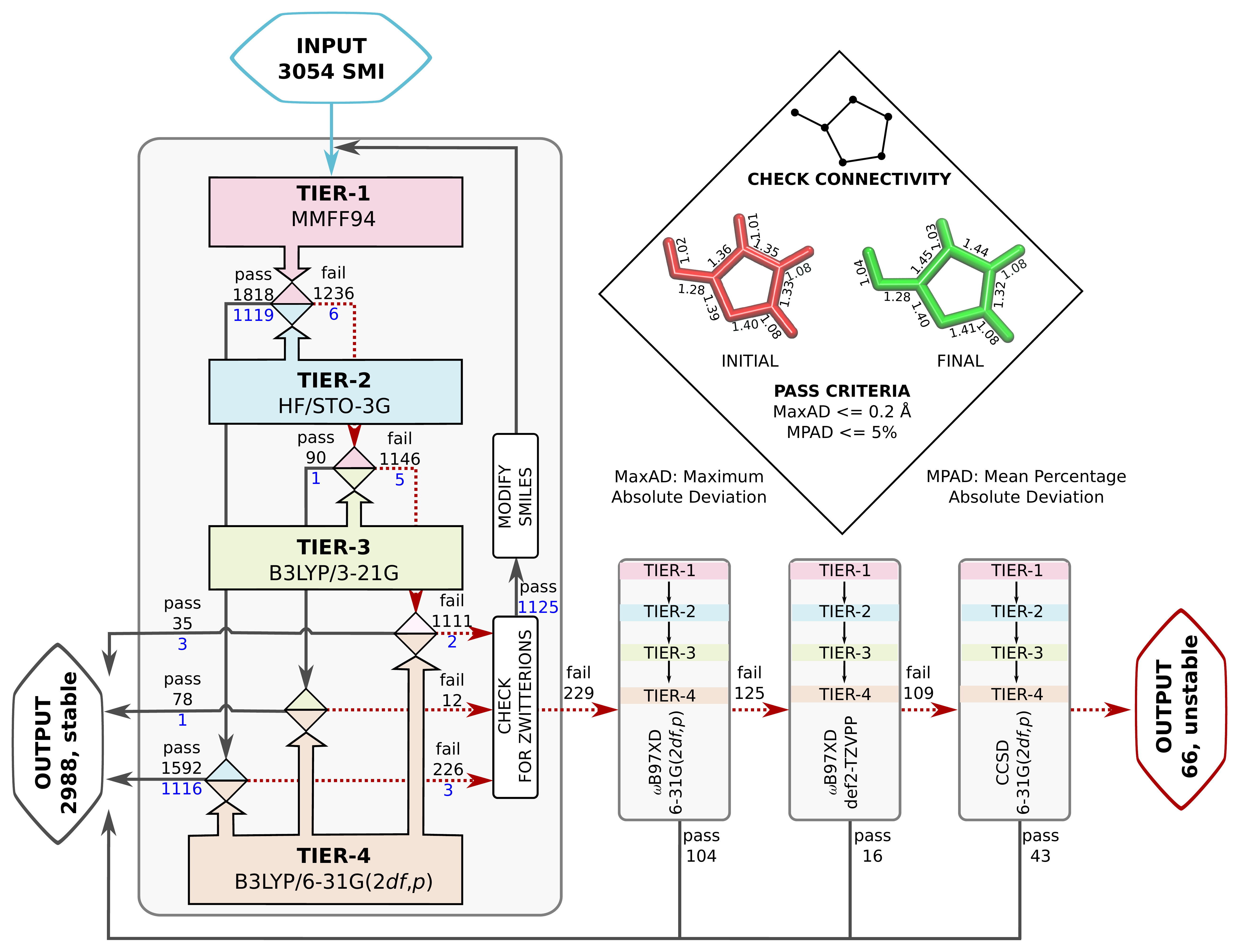}
    \caption{
    Automated workflow for 
    {\bf conn}ectivity preserving {\bf g}eometry {\bf o}ptimizations ({\tt ConnGO}).
    The scheme is illustrated for the 3k set
    with pass/fail
    statistics. 
    Geometries from hierarchically improved methods (tiers) are
    checked for connectivity conservation
    using the criteria highlighted in the inset (top, right).
    At every tier, the final minimum energy geometry is compared
    with the initial one. 
    {\tt ConnGO} takes 3,054 SMI as input and converges 
    2,825 as local minima when the tier-4 is B3LYP/6-31G(2{\it df},{\it p}),
    while 163 molecules are subsequently stabilized with better tier-4 options.
    }
    \label{fig:conngo}
\end{figure*}
\begin{figure*}[ht]
    \centering
    \includegraphics[width=13cm]{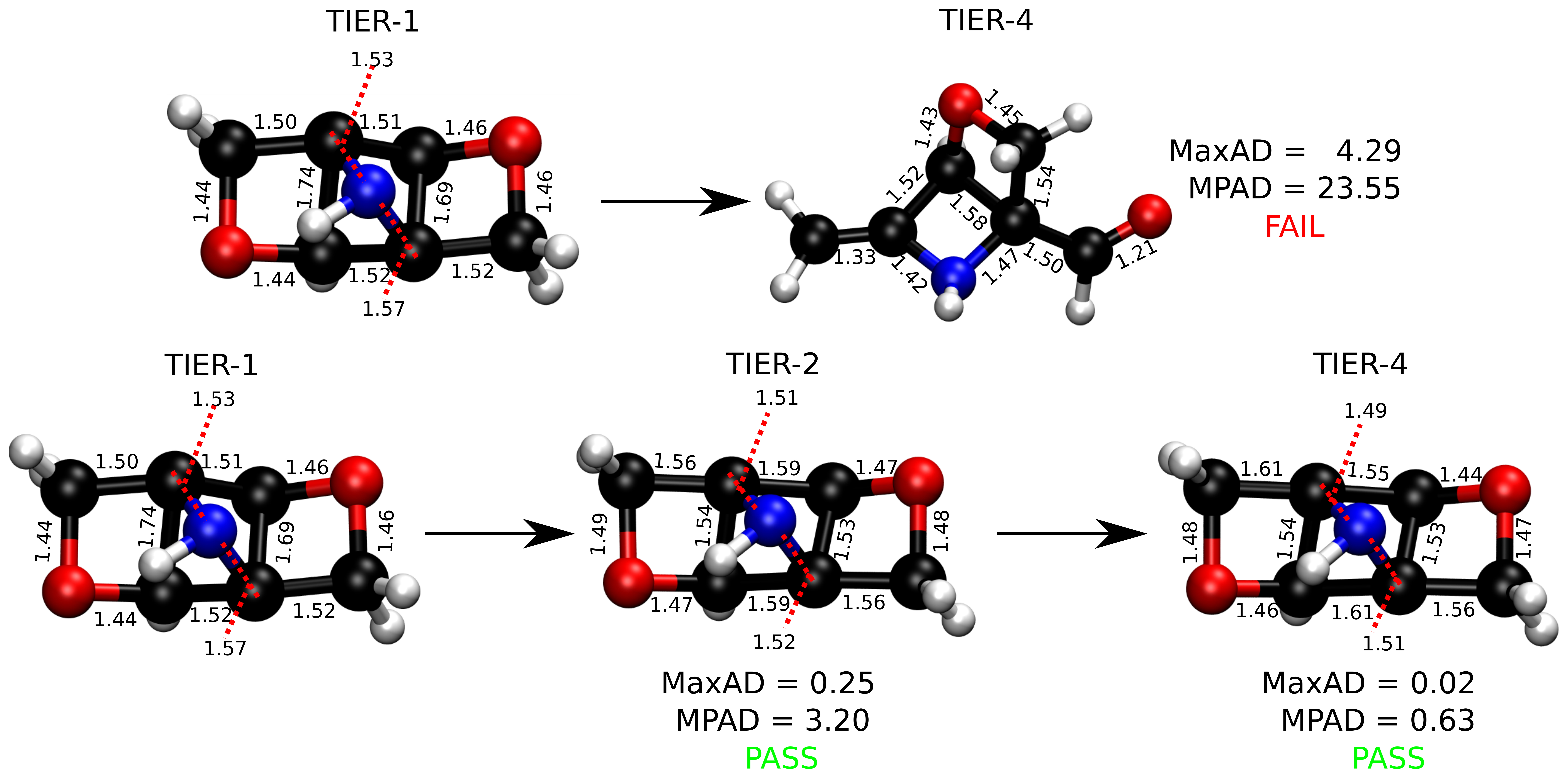}
    \caption{An example molecule from the 3k set undergoing rearrangement during
    DFT optimization while starting with a force field geometry. Successful
    optimization of the same molecule with the {\tt ConnGO} workflow using
    an intermediate tier is also illustrated. Bond lengths and
    MaxAD are in \AA.}
    \label{fig:selectflows}
\end{figure*}



In the present study, we propose and benchmark an automated workflow for 
{\bf conn}ectivity preserving {\bf g}eometry {\bf o}ptimizations ({\tt ConnGO}) 
using multiple tiers of quantum mechanical approximations for 
high-throughput geometry optimizations.
The proposed approach, illustrated in 
Fig.~\ref{fig:conngo}, was tested 
for a multitude of control parameters---the choice of force fields 
({\bf Figures~S1 \& S2}), optimizers ({\bf Figure S3}), and  
combinations of methodological tiers ({\bf Figure S4}) 
---and only the best performing set up is discussed here. See Supplementary Information (SI) for more details.
{\tt ConnGO} takes as input a file containing SMI that are  
converted to 3D Cartesian coordinates in the SDF format using Openbabel~2.3.2\cite{o2011open}. 
These initial coordinates are relaxed with the Merck molecular force field (MMFF94)\cite{halgren1996merck} via steepest descent minimizer enforcing a tight threshold of $10^{-8}$ kcal/mol for energy convergence. 
Henceforth, we denote these settings as tier-1 in {\tt ConnGO}; in {\bf Figure~S3} we show this set up to yield superior performance over other variations. Geometries
relaxed at tier-1 are stored in the SDF format, with 
the same bonding connectivities encoded in the corresponding SMI.
All tier-1 geometries are further relaxed at tier-2; 
for this purpose, Hartree--Fock (HF) with a minimal basis set has 
been identified as suitable among several methods (see {\bf Figure~S4}).
To probe for geometric rearrangements or dissociation, we use 2 metrics:
the maximum absolute deviation (MaxAD) of bond lengths corresponding to covalent connectivities 
and their mean percentage absolute deviation (MPAD). 
In order not to rule out the possibility of \textit{unusual} structures, those with
 MPAD $<5\%$ and MaxAD $>0.2$~\AA{} are deemed {\it pass} in tier-2 if their tier-1 geometries have bond lengths $>1.70$~\AA{}.

Geometries converging at tier-2---fulfilling tier-1 vs tier-2 pass criteria---are sent directly to tier-4 with the target DFT-level, B3LYP/6-31G(2{\it df},{\it p}), while those that {\rm fail} enter the intermediate
tier-3, B3LYP/3-21G, starting with the tier-1 geometries. 
Final optimized geometries from each tier are 
compared with those from the previous tier and
checked for changes in connectivities. 
In addition, geometries are verified at tiers-2/3/4
for being a local minimum on the 
potential energy hypersurface through vibrational analysis. 
SMI of molecules failing tier-4 are checked for 
zwitterionic character; when detected, such SMI are converted to 
their neutral forms that enter another iteration of {\tt ConnGO}.
In view of potential drawbacks of the B3LYP method to treat complex
bonding situations, molecules that fail tier-4 are subjected to 
{\tt ConnGO} based on the improved tier-4 settings: 
$\omega$B97XD/6-31G(2{\it df},{\it p}),
$\omega$B97XD/def2-TZVPP and
CCSD/6-31G(2{\it df},{\it p}). All tier-2/3/4
calculations were performed
using Gaussian16 suite of programs\cite{frisch2016gaussian}.
The documentation and examples collected at 
\href{https://github.com/salinisenthil/ConnGO}{{\tt https://github.com/salinisenthil/ConnGO}}
 show how {\tt ConnGO} can be modified to use with other software packages and methods beyond those utilized in the present work.


For the stable 131k set, we generated 
tier-1 geometries that in comparison with previously reported\cite{ramakrishnan2014quantum} 
DFT-level geometries \textit{pass} {\tt ConnGO} criteria. 
Hence, in this study, we only focus on the remaining 3k set.
To maintain uniformity in the quality of the 3k final 
geometries with that of the 131k set studied before, we
use the same DFT settings for tier-4.
The preference for this theory was
motivated by the fact that the G$n$ series of  
composite methods\cite{curtiss2011gn} depend on minimum energy
geometries and scaled harmonic
frequencies at this very level.
\begin{figure*}[ht]
    \centering
    \includegraphics[width=17.0cm]{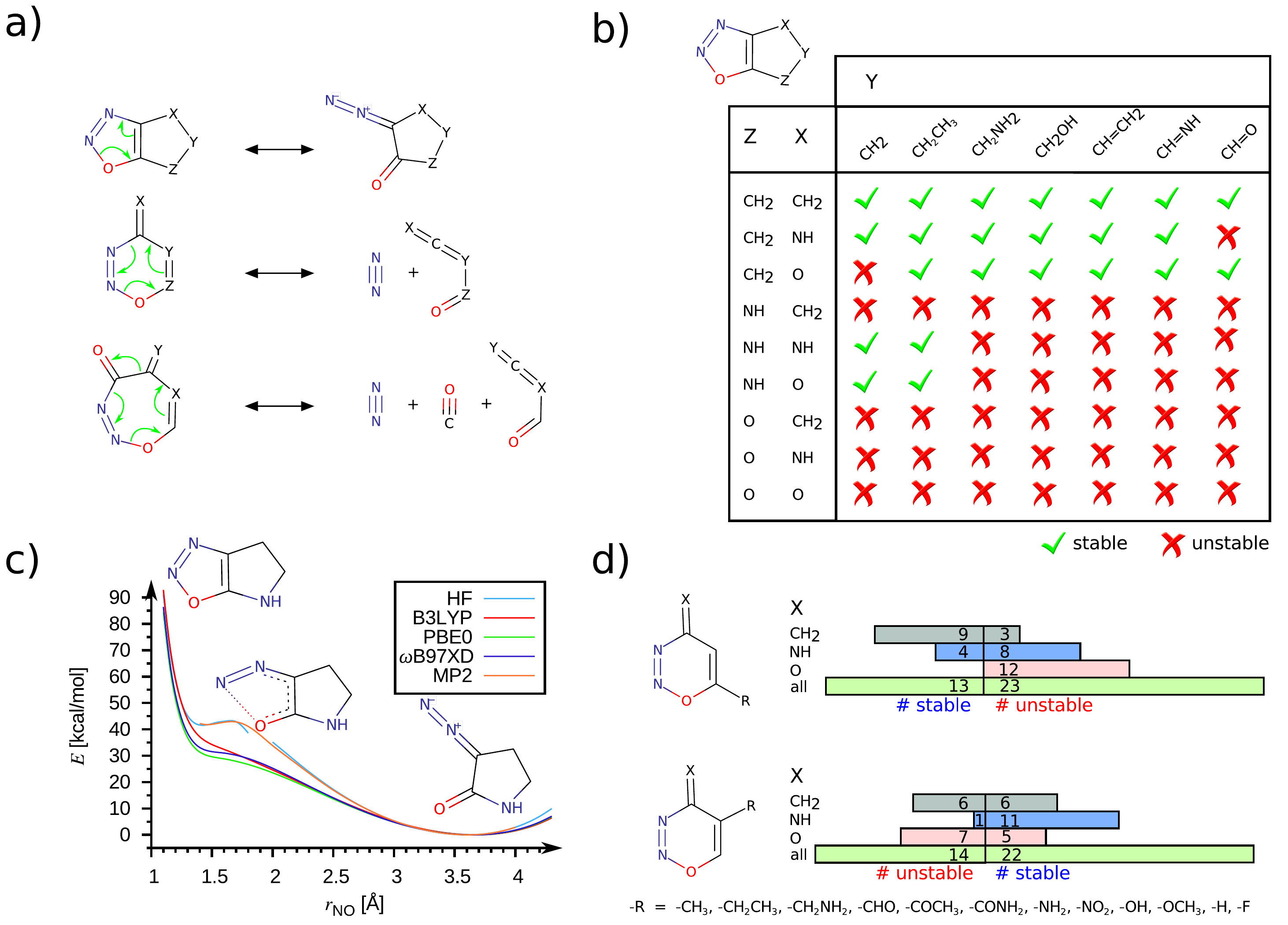}
    \caption{Stability trends across  $-{\rm NNO}-$ 
    containing heterocycles:
    a) Mechanistic pathways for ring-opening or fragmentation is
    shown for 5/6/7-membered ring systems. 
    b) Effect of regio-selective substitutions 
    on the dynamic stability of 
    5-membered heterocycles. During geometry optimizations, the unstable structures (red crosses) 
    undergo valence-tautomeric rearrangements to form the diazo-keto product.
    c) Ring-opening reaction path for an example
    5-membered system. For a given 
    $r_{\rm NO}$, all other internal coordinates are 
    relaxed and the corresponding potential energies ($E$)
    are plotted for various
    levels of theory.
    d) Effect of regio-selective substitutions 
    on the dynamic stability of 
    6-membered heterocycles.
    All results are from $\omega$B97XD/def2-TZVPP level unless
     stated otherwise.
    }
    \label{fig:NNO}
\end{figure*}
\begin{figure*}[ht]
    \centering
    \includegraphics[width=18.0cm]{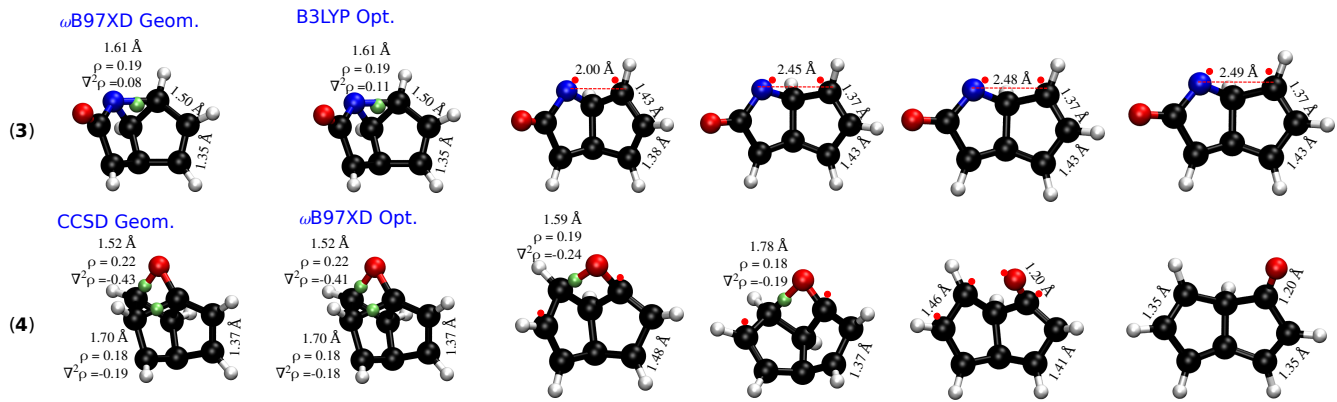}
    \caption{
   Dissociation trajectories of selected molecules during optimization at a lower level where the initial structures are
   higher-level minima. 
     $-{\rm NNO}-$  containing molecules (top) and highly strained molecules (bottom): 
    ({\bf 1}) \& ({\bf 3}) are $\omega$B97XD minima dissociating at the 
    B3LYP level; 
    ({\bf 2}) \& ({\bf 4}) are CCSD minima
    dissociating at the $\omega$B97XD level.
    BCPs are marked by green spheres; $\rho$ and $\nabla^2 \rho$ at BCPs are reported in $e$/bohr$^3$ and $e$/bohr$^5$, respectively.
    Red dots denote radical sites.
    All results are obtained with the 6-31G(2{\it df},{\it p}) basis set.
    }
    \label{fig:nnodscheme}
\end{figure*}
To illustrate the need for iterative geometry optimizations for
reliable high-throughput generation of molecular structures,
we identified molecules for which a direct target
tier-4 optimization
starting with tier-1 geometry failed to converge. 
This is often the case when equilibrium structures at both levels
differ substantially, as illustrated in Fig.~\ref{fig:selectflows}. 
Such situations benefit from the 
hierarchical treatment in {\tt ConnGO} that exploits the
nearness in the theoretical models and 
introduces levels of intermediate rigour. 
The cornerstones of quantum chemistry modelling, 
Pople diagrams\cite{pople1999nobel} and Perdew's Jacob's ladder\cite{perdew2001jacob}, have also
benefited from the {\it nearness} heuristic.

Upon analysis, we found 1,262 entries 
of the 3k set to be zwitterions, of which
1,125 failed in the first round of 
the {\tt ConnGO} flow (see Fig.~\ref{fig:conngo}).
These molecules re-entered the workflow, starting with the corresponding neutral SMI;
pass/fail statistics are reported in {\bf Figure~S5}.
Overall, the methodology proposed in this study decreases
the number of equivocal molecular cases from 3,054 to 
229. This number is further decreased to 66 by
repeating {\tt ConnGO}-based iterative geometry
optimizations with accurate 
$\omega$B97XD and 
CCSD levels. 
To facilitate future explorations, SMI of the unstable set are provided in {\bf List S1}. 
Among the 66 dynamically unstable molecules, 52 are heterocycles containing 
an  $-{\rm NNO}-$  moiety. Further analysis  
identifies these to have 5-, 6-, or 7-membered rings 
that dissociate via the pathways illustrated 
in Fig.~\ref{fig:NNO}a. 
In all cases, the final
structure displays a broken NO bond; their
lability must arise from a subtle combination of
the ring structure and substitutions as there are
185 stable $-{\rm NNO}-$ systems in the 2,988 newly characterized set featuring
conventional 
NO distances (see {\bf Figure~S6}).

It has been previously noted that the simplest $-{\rm NNO}-$ 5-ring, 1,2,3-oxadiazole, 
undergoes ring-opening to form its diazo-keto 
valence-tautomer\cite{joule2008heterocyclic}. 
An earlier work\cite{nguyen1985can} based on a small basis set  predicted
1,2,3-oxadiazole to be unstable in the gas phase.
The same molecule fused with a benzene ring, 1,2,3-benzoxadiazole,
has been characterized in an argon matrix using infrared (IR) spectroscopy at 15 K\cite{schulz19841}. In {\bf Figure S7}, 
using DFT methods we show both these compounds to be dynamically
stable. In this study, we identify several bicyclic 5-ring compounds to be
dynamically unstable, resulting in
NO $\sigma$-bond cleavage. In 
Fig.~\ref{fig:NNO}, we present a consolidated stability map of combinatorially
generated
$-{\rm NNO}-$ containing 5/6-rings.
While only some of the molecules presented in this analysis 
feature in the QM9 set, the others were generated afresh for a
 complete coverage.
We find the combination of functional groups at X, Y and Z positions
to modulate the stability of the bicyclic rings; the unstable ones
undergo ring-opening 
valence-tautomerism to stabilize 
the diazo-keto isomer (see Fig.~\ref{fig:NNO}a, b). 
When the group at the Z-position is 
-CH$_2$-, which is inductively electron-donating (+I), 
the$-{\rm NNO}-$ring is stabilized, Fig.~\ref{fig:NNO}b. 
Replacing the methylene group with isoelectronic variants
-NH- or -O-, containing electronegative centers, promotes ring opening. 
When the Z group is 
oxo, the choices at X and Y have no effect. 
The trends noted 
here are mostly independent of the DFT methods. 
To demonstrate this, we have plotted the reaction path for ring-opening
modelled at the HF, MP2 levels and with selected DFT methods 
in Fig.~\ref{fig:NNO}c. 
Both HF and MP2 metastable profiles
show a small barrier ($<3$ kcal/mol) at $r_{\rm NO}\approx1.70$~\AA{}
preventing the molecule
to escape to its dissociated form. 
On the other hand, DFT methods
show a vanishing curvature at
$r_{\rm NO}\approx1.40$~\AA{} indicating an inadequate well-depth
to host vibrational bound states for the ring structure. 
More conclusive characterization of the stability will require post-MP2 level treatments
that become prohibitively expensive for high-throughput explorations.

The 6/7-ring
compounds undergo  
electrocyclic ring-opening to 
dissociate into dienes and dienophiles (Fig.~\ref{fig:NNO}a).
The regio-selective stabilization of the 6-ring as 
a function of X, Y and Z groups in the 
ring is illustrated in Fig.~\ref{fig:NNO}d. For fixed X and Z,
we probed how the modulation of the ring current by +I
groups at Y- and Z-terminals will facilitate the 
retro Diels-Alder reaction. Of the 
36 compounds selected with variations 
at Y-position, we find 14 to be unstable
following a dissociation path. Whereas for variations made at the 
Z-position, the number of stable compounds 
decreased indicating perturbation in the environment of$-{\rm NNO}-$group's O atom 
to have a role in 
diminishing the stability of the 6-ring. Of the 52 unstable
$-{\rm NNO}-$ systems, 3 are 7-rings dissociating into diene, N$_2$ and CO 
molecules as shown in Fig.~\ref{fig:NNO}a.

To understand the dependence of the covalent bonding lability on the rigour of the
theoretical method employed, we performed 
topological analysis of electron density\cite{bader1981topological}  using the Multiwfn software\cite{lu2012multiwfn}.
Fig.~\ref{fig:nnodscheme} presents the cases of
2  $-{\rm NNO}-$  containing rings:
4H-1,2,3-oxadiazin-4-one ({\bf 1}) and
4H-1,2,3,5-oxatriazin-4-imine ({\bf 2});
and 2 highly-strained molecules:
2-azatricyclo[3.3.0.0$^{2,8}$]octa-4,6-dien-3-one ({\bf 3}) and
2-oxatetracyclo[4.2.1.0$^{1,4}$.0$^{3,9}$]nona-5,7-diene ({\bf 4}). 
In each category, one of the molecules is stabilized at the CCSD level, but
undergoing ring-cleavage at the DFT-level. 
The other molecule is 
stable at the $\omega$B97XD level, while unstable at the B3LYP level. Although it has been argued that the existence of a 
bond critical point (BCP) is not direct evidence of bonding\cite{shahbazian2018bond,wick2018bond},
qualitative conclusions can be drawn from the
electron density ($\rho$) and its Laplacian ($\nabla^2 \rho$) at the BCP.

Molecule ({\bf 1}) when relaxed at 
the $\omega$B97XD-level exhibits a stable structure.
The local minimum from the $\omega$B97XD level, when further 
relaxed with B3LYP, fragments to 3-oxoprop-2-enal and N$_2$. 
Similarly, ({\bf 2}) that is stable at CCSD, fragments to N-carboximidoylformamide and N$_2$ when treated with $\omega$B97XD. 
The dissociation of both molecules follows the pathway shown 
in (Fig.~\ref{fig:NNO}a), highlighting 3 bond-breaking and 3 bond-making processes.
Molecule ({\bf 1}), at the $\omega$B97XD geometry shows quantitatively
similar $\rho$ and $\nabla^2 \rho$ at B3LYP
and $\omega$B97XD, indicating the electron density distribution to be very
similar; nevertheless the structure is a local
minimum in only one of the methods. B3LYP relaxation progresses with a 
gradual accumulation of $\rho$ along the NN bond, which is also
reflected by the corresponding $\nabla^2 \rho$ becoming more negative. 
In contrast, $\rho$ and $\nabla^2 \rho$ of 
the NO and CN bonds consistently drop during relaxation, finally 
vanishing upon fragmentation.

 \begin{figure}[h]
    \centering
    \includegraphics[width=8.5cm]{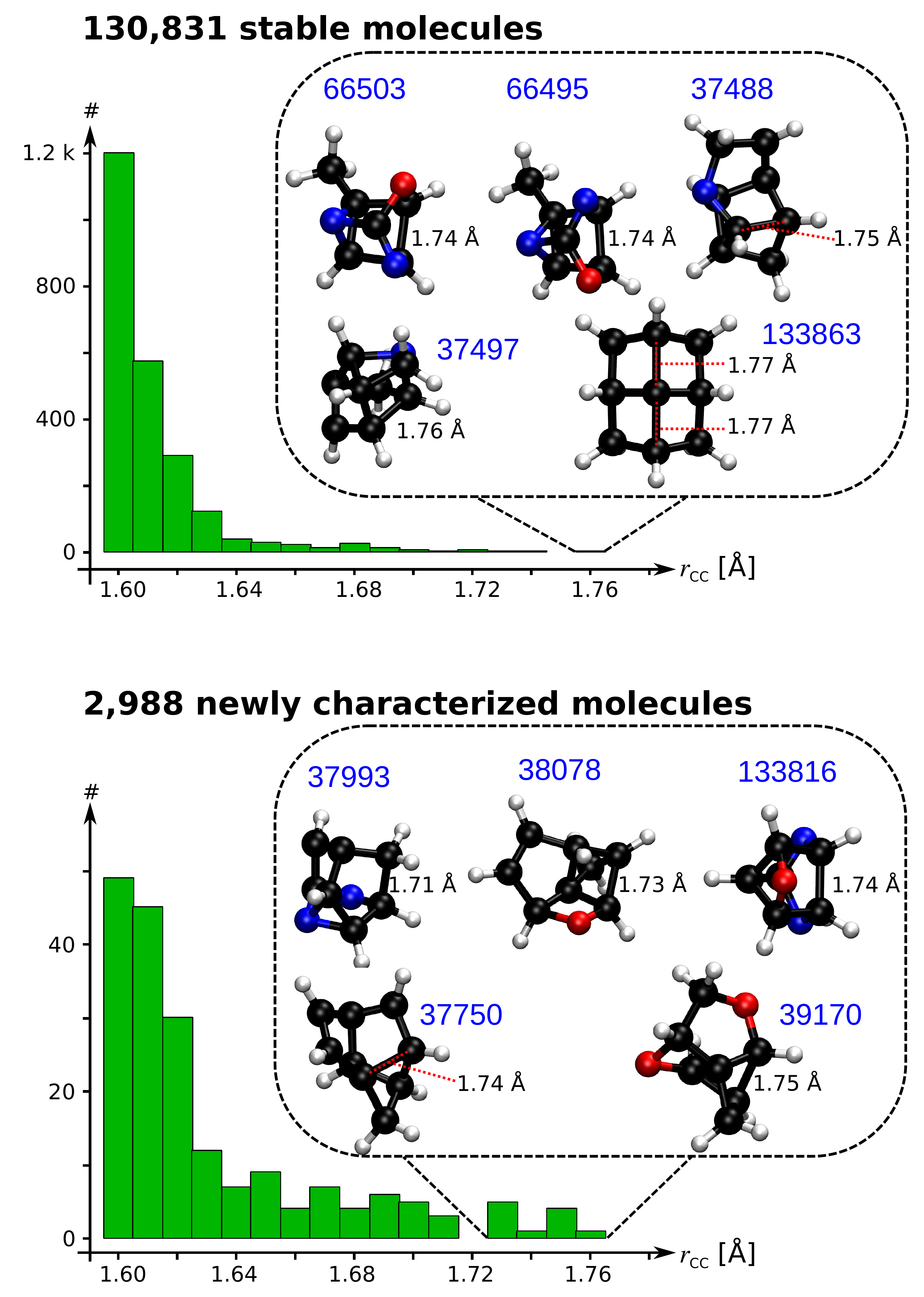}
    \caption{Distribution of long CC bond lengths ($r_{\rm CC}$) in the 131k set (top) and 
    curated 3k set (bottom).
    The insets show candidate molecules with ultralong 
    $r_{\rm CC}$ with their QM9 indices in blue. 
    }
    \label{fig:longbond}
\end{figure}
Bond dissociation in the strained cases ({\bf 3}) and ({\bf 4}) are 
marked by unusually long single bonds that are adequately treated only
when the theoretical level captures long-range effects of electron correlation.
Structural rearrangements in ({\bf 3}) and ({\bf 4}) are driven by 
long CN ($r=1.61$~\AA{}) and CC bonds ($r=1.70$~\AA{}), respectively.
In both cases, the end product is a biradical stabilized by an
allylic conjugation.
Topological analysis indicates an increase in $\nabla^2 \rho$ to
mark the onset of structural rearrangements.
While it may be argued that the 
unfavourable description of the electron distribution, $\nabla^2 \rho \ge 0$, at the BCP of a covalent bond---NO in ({\bf 1}) and ({\bf 2}); CN/CC in ({\bf 3}) and ({\bf 4})---is a factor
driving the dissociation, based on the evidence tendered above their close dependence on the theoretical level remains vague.

As a data-mining exercise to reveal structural diversity
across the curated QM9
dataset,
we search for candidate molecules with ultralong CC single bonds ($r_{\rm CC}>1.70$~\AA{}).
Fig.~\ref{fig:longbond} presents the distributions of CC bond lengths
for $r_{\rm CC}>1.60$~\AA{};
results for the entire range are presented in {\bf Figures S8 \& S9}.  
Even though the search for long CC bonds is not new,  the principles 
guiding the detection of such evasive bonding remains largely unknown\cite{laura2019the}. 
One of the longest alkane bond lengths ever reported (1.704~\AA) is 
obtained by coupling a diamantane with a triamantane\cite{schreiner2011overcoming}. While other reports have
suggested longer bond lengths, the exact classification of their characteristics---based on the fraction of charge transfer/ionic/biradical interactions---requires further investigations. 
Fig.~\ref{fig:longbond} illustrates $r_{\rm CC}>1.60$~\AA{} and highlights the Top-10 candidates with a BCP; their geometries are collected in {\bf List S2}. The structural variation across the systems show that accruing a long CC bond is system-specific and requires a careful balance between stabilizing and destabilizing interactions within a molecule. Remarkably, all systems showcased in Fig.~\ref{fig:longbond} contain at most 9 heavy atoms compared to those identified in previous studies, which contain several main group atoms. Furthermore, we identified 8 molecules
in the curated 3k set with $r_{\rm CC}>1.90$~\AA{} (not shown in Fig.~\ref{fig:longbond}). Due to the absence of a BCP along these long CC distances, we
believe the corresponding molecules to be of biradical nature with a reduced bond order.


\begin{table}[h]
\small
    \caption{Long CC bonds (in \AA) for the Top-10 molecules highlighted in Fig.~\ref{fig:longbond}, computed at B3LYP and $\omega$B97XD methods with the def2-TZVPP basis set. Also stated in parenthesis are  $\rho$ (in $e$/bohr$^3$) and $\nabla^2 \rho$ ($e$/bohr$^5$) values at the bond critical points. Missing values indicate dissociating molecules. } 
    \centering
    \begin{tabular}{rrrrr}
    \hline 
\multicolumn{1}{l}{Index}  & & \multicolumn{3}{l}{$r_{\rm CC}$ ($\rho$, $\nabla^2 \rho$) }  \\
\cline{3-5}
&  &  \multicolumn{1}{l}{B3LYP} &&
\multicolumn{1}{l}{$\omega$B97XD}  \\
          \hline
37488 &~~&      1.750 (0.161, -0.200) &~~&      1.700 (0.178, -0.268) \\
37497 &&        1.783 (0.153, -0.187) &&        1.712 (0.176, -0.277) \\
37750 &&                                  &&    1.757 (0.155, -0.156) \\
37993 &&        1.716 (0.169, -0.239) &&        1.692 (0.177, -0.270) \\
38078 &&        1.746 (0.158, -0.169) &&        1.703 (0.173, -0.228) \\
39170 &&                                  &&    1.751 (0.161, -0.188) \\
66495 &&        1.734 (0.174, -0.249) &&        1.705 (0.183, -0.289) \\
66503 &&        1.732 (0.174, -0.253) &&        1.704 (0.184, -0.292) \\
133816 && 1.737 (0.256, -0.585) && 1.699 (0.185, -0.300) \\
133863 && 1.797 (0.130, -0.077) && 1.735 (0.150, -0.145) \\
        \hline
    \end{tabular}
    \label{tab:longbondCC}
\end{table}

The longest bond length in Fig.~\ref{fig:longbond} is noted for the 
pyramidal conformer (point group, C$_{4v}$) of fenestrane, at the B3LYP/6-31G(2{\it df},{\it p}) level, with $r_{\rm CC}=1.77$~\AA{} agreeing with a previous report\cite{schulman1983structure}. 
The same study also reported the CC distance 
dropping to $r_{\rm CC}=1.49$~\AA{} for the
competing pyramidal conformer (point 
group, D$_{2d}$)---establishing the thermodynamic
stability of these conformers goes beyond the scope of the present study. 
In the curated 3k set, 
the top candidate is 5,8-dioxatricyclo[4.3.0.0$^{3,9}$]nonane with $r_{\rm CC}=1.75$ \AA. To understand the sensitivity of $r_{\rm CC}$ to the theory, 
we studied the Top-10 cases with B3LYP and $\omega$B97XD using a large basis set; SMI and results from other methods are collected in
{\bf Table~S1}. 
The corresponding bond lengths along with  
$\rho$ and $\nabla^2 \rho$ calculated at the BCPs are collected in
Table.~\ref{tab:longbondCC}. In general, bond lengths predicted with B3LYP 
are overestimated compared to the $\omega$B97XD values. This trend is also reflected by the diminished values of $\rho$ and its $\nabla^2 \rho$ with B3LYP.
Nevertheless, at the more accurate $\omega$B97XD, a number of molecules
exhibit $r_{\rm CC}>1.70$~\AA{} warranting further investigations 
using more rigorous methods for quantitative modelling of
vibrational spectra, thermodynamic stability and competing 
rearrangement reaction
pathways, suggesting directions towards experimental detection.


In conclusion, an iterative scheme for 
automated geometry optimization 
addresses the difficulties associated with conserving covalent bonding
connectivities in high-throughput chemical space explorations. 
Out of 3,054 molecules with structural ambiguities, we curated
2,988 using the workflow proposed in this study; the remaining
molecules are deemed dynamically unstable.
It is important to note that a conclusive diagnosis of the unstable
molecules as  $-{\rm NNO}-$  heterocycles or 
those with pyramidal sp$^2$ C---in combination with their unique chemical environment---requires an accuracy better than DFT. 
The unique balance between stabilizing and destabilizing bonding factors
makes the  $-{\rm NNO}-$  containing 6- or 7-ring metastable molecules
potential candidates for 
high-energy materials\cite{he2015bicyclic}; the 6-ring maybe
preferred---for, its only side product is N$_2$. 
Future studies can also tailor small organic molecule with ultralong CC bond, 
through rational modifications of the systems
presented in this work. 
While a universal
approach to characterize, {\it a priori}, whether a molecular structure will dissociate or rearrange is still lacking, automated 
determination of BCP characteristics at the initial-guess geometry
could provide a remedy incurring additional computational costs. 
Further benchmarking of the strategy 
proposed here can also begin with force fields containing anharmonic terms\cite{allinger2016molecular,shannon2019anharmonic},
suitable for describing unusually long covalent bonds. 
Big data initiatives must pay close attention to molecules
failing to converge and inspect their properties rather than 
discarding them---for, the factors behind their unusual
behaviour may stem from prospective chemical trends as of yet unexplored.


\section*{Acknowledgements}
SS is grateful to TIFR for Visiting Students’ Research Programme (VSRP)
and junior research fellowships.
This project was funded by intramural funds at TIFR Hyderabad from 
the Department of Atomic Energy (DAE).
All calculations have been performed using the Helios computer cluster, 
which is an integral part of the MolDis Big Data facility, 
TIFR Hyderabad 
(\href{https://moldis.tifrh.res.in/}{\tt https://moldis.tifrh.res.in/}).





\bibliography{rsc} 

\end{document}